\documentstyle[aps,prb,multicol,graphics,epsf]{revtex}
\begin{document}
\draft

\title{Hubbard model with orbital degeneracy and integer or noninteger    
filling}                 
\author{O. Gunnarsson}
\address{Max-Planck-Institut f\"ur Festk\"orperforschung,
         D-70506 Stuttgart, Germany}
\date{\today}

\maketitle

\begin{abstract}
The photoemission spectrum and the specific heat are studied
in the Hubbard model of small clusters including orbital degeneracy. 
We focus on the degeneracy and valence dependence in the limit of
 a large Coulomb interaction. For integer valence, it is found that 
the degeneracy increases the width of the photoemission spectrum
and it reduces the contribution from the charge degrees of freedom
to the specific heat. A deviation from integer valence reduces the
width of the photoemission spectrum.                       
\end{abstract}
\pacs{71.10.Fd,71.20.+h}

\begin{multicols}{2}
\section{Introduction}
Ni metal provides a typical example of the photoemission spectrum
from a relatively strongly correlated system.\cite{Huefner}
The width of the Ni $3d$ band\cite{Nipes} is reduced by about 25 $\%$ 
compared with one-particle (density functional) calculations.\cite{Wang}
As a further signature of many-body effects, the spectrum has a satellite
(``6 eV satellite''),\cite{Nipes} which is displaced  by roughly the
Coulomb interaction $U$ between two $3d$ electrons on a 
Ni atom.\cite{Nicalc}
In Ni the many-body effects are primarily due to the localized nature of the
$3d$ orbital. The $3d$ band has a noninteger occupation ($\sim 0.6$ holes), since the $4s$ orbital is also partly occupied.

 A different example of a rather strongly correlated system 
is provided by alkali-doped C$_{60}$ compounds, A$_3$C$_{60}$
(A= K, Rb). It is believed that the alkali atoms are almost 
completely ionized,\cite{Martins,Satpathy} and the three alkali
electrons are therefore donated into  the  $t_{1u}$ orbital of 
the C$_{60}$ molecule, which has a threefold orbital degeneracy. 
Since in this case        
essentially only one orbital ($t_{1u}$) is partly occupied, 
it has an almost integer occupancy. These systems have slight 
deviations from stoichiometry, but the occupancy should nevertheless
be almost integer.
This is in contrast to the noninteger 
occupancy of the local orbital in many other correlated systems.

For K$_3$C$_{60}$ it has not been possible to determine the dispersion
of the bands using angular resolved photoemission, due to the presence 
of orientational disorder and problems related to the angular resolution
in relation to the small size of the Brillouin zone. Measurements of
the specific heat show, however, that there is no enhancement of
the specific heat beyond what would be expected from the electron-phonon
interaction.\cite{Meingast,susc} This is quite surprising in 
view of the expected strong correlation effects in these systems.

The on-site Coulomb energy $U$ between two electrons
on the same C$_{60}$  molecule is large\cite{Lof} compared with the 
one-electron width $W$ of the $t_{1u}$ band, with 
$U/W\sim 1.5-2.5$.\cite{MottC60,RMP} For such a large ratio one 
might have expected the system to be a Mott-Hubbard 
insulator.\cite{Georges} We have shown, however, that at half-filling 
the ratio of $U/W$ where the transition takes place  grows with
the orbital degeneracy $N$ of the $t_{1u}$ orbital as $\sqrt{N}$ or somewhat
faster.\cite{MottC60,LargeU} In the large $U$ limit it was shown
that an extra electron in a system with otherwise integer occupancy,
e.g., half-filled,
can move more efficiently if the orbital degeneracy $N$ is large. 
The reason is that the extra occupancy can move through a hop
of any of the electrons on the site with the extra 
occupancy.\cite{MottC60,LargeU} In the half-filled case, of the order
$N$ electrons can hop to a neighboring site. 
For a different integer filling than half-filling, this effect 
is reduced.\cite{filling}

We observe that in a strongly correlated system hopping is in general
reduced compared with the corresponding noninteracting system. 
Due to the arguments given above,
the degeneracy nevertheless enhances the effects of hopping
on the Mott transition compared with what might have been
expected based on the importance of hopping for the noninteracting
system. It is then interesting to ask if this may also happen
for other properties. 

Here we study photoemission spectroscopy (PES) for the Hubbard
model of some finite systems. We show that for these systems 
and integer occupancy, the width of the PES spectrum is larger
than the noninteracting band width $W$ for a large $U$ and $N>1$,
except for filling one and $2N$.
These are therefore  examples where correlation effects {\sl increase}
the band width.
The width is reduced as $U$ is reduced or as the valence starts
to deviate from an integer and may then become smaller than $W$.
Such systems with noninteger occupation are therefore cases
where correlation effects reduce the band width.
Systems with noninteger occupancy also develop a satellite which moves 
roughly linearly with the value of $U$, while such satellites have very 
small weights for systems with integer occupancy.
We have given
a brief discussion of the angular integrated photoemission spectrum 
for a system with integer valence before.\cite{Antropov}

In addition we discuss the contribution to the specific heat 
from charge degrees of freedom. We show that the orbital degeneracy
can reduce this contribution, which may be the reason for the         
lack of an electronic enhancement of the specific heat in K$_3$C$_{60}$.

After giving the formalism (II.A) and presenting some general
arguments (II.B), we show
results for the photoemission spectrum for a diatomic
molecule (Sec. II.C), a linear chain (Sec. II.D),     a six atom
cluster (Sec. II.E),  a four molecule model of K$_3$C$_{60}$ 
(Sec. II.F) and with multiplet effects (Sec. II.G). The specific heat
is treated in Sec. III and the results are discussed in Sec. IV. 

\section{Photoemission spectrum}

\subsection{Formalism}
Below we study multi-band Hubbard models 
\begin{eqnarray}\label{eq:1}
  H &&= \sum_{<ij>m\sigma} t_{im,jm^{'}}\,\psi^{\dagger}_{im\sigma}
\psi_ {jm'\sigma}  \nonumber  \\
   && + U\sum_i\sum_{(m \sigma) < (m'\sigma')}n_{im\sigma}n_{im'\sigma'},
\end{eqnarray}
where $i$, $m$ and $\sigma$ are site, orbital and spin indices, 
respectively. The orbital index can take $N$ values. 
 The hopping integrals are given by $t_{im,jm^{'}}$
and $U$ is the on-site Coulomb integral. 
All multiplet effects have been neglected, except in Sec. II.G.
The creation operator
for an electron with the quantum numbers $i$, $m$ and $\sigma$ is 
given by $\psi^{\dagger}_{im\sigma}$.          
We are interested  in the spectrum for removing an electron 
with a given quantum number $k$
\begin{equation}\label{eq:2}
\psi_k=\sum_{im\sigma}c(k)_{im\sigma}\psi_{im\sigma}.
\end{equation}
The spectrum is then given by
\begin{eqnarray}\label{eq:3}
&&\rho_k(\varepsilon)=  \\
&&\sum_n|\langle M-1,n|\psi_k|M,0\rangle|^2
\delta(\varepsilon-E_0(M)+E_n(M-1)), \nonumber
\end{eqnarray}
where $|M,0\rangle$ is the initial ground-state with $M$ 
electrons and $|M-1,n\rangle$ is an excited final state. 
The corresponding energies are $E_0(M)$ and $E_n(M-1)$, 
respectively. We use Lanczos method to calculate the 
ground-state and the spectrum.\cite{Lanczos}
In most, but not all, cases
we assume that the hopping matrix elements are ``diagonal''
in $m$ and $m^{'}$
\begin{equation}\label{eq:4}
t_{im,jm^{'}}=\cases{t\delta_{mm'},& if $i$ and $j$ nearest neighbors;\cr
                     0,& otherwise.\cr}
\end{equation}
Here we assume $t$ to be negative.

\subsection{General considerations}

We first address the hopping of an electron or a hole in a system
which otherwise has integer occupancy. We consider the large $U$-limit.
We construct a Neel state $|0\rangle$ with $N$ electrons of 
the same spin per site. We now add an extra electron to site 1
and obtain the state 
\begin{equation}\label{eq:a}
|1\rangle=\psi^{\dagger}_{11\downarrow}|0\rangle.
\end{equation}
The extra occupancy on site 1 can now hop to the neighboring
sites $i$ without any extra cost in Coulomb energy. We can form a
state where the extra occupancy is on the neighboring site $i$
\begin{equation}\label{eq:b}
|i\rangle={1\over \sqrt{N}}\psi^{\dagger}_{11\downarrow}
\sum_m\psi^{\dagger}_{im\uparrow}
\psi_{1m\uparrow}|0\rangle.  
\end{equation}
We have here taken into account that any of the $N$ spin up electrons
on site 1 can hop to site $i$, as is illustrated in Fig. \ref{fig0}a.
Assuming hopping matrix elements of the form (\ref{eq:4}), we obtain
the matrix element\cite{MottC60}
\begin{equation}\label{eq:c}
\langle i|H|1\rangle=\sqrt{N}t,
\end{equation}
where $t$ is the one-particle matrix element. Thus the hopping matrix element 
has been enhanced by a factor of $\sqrt{N}$.
In this argument we have assumed that the electron hops against a
background of a classical Neel state. For certain systems, we have 
constructed states where the hopping energy of the extra electron or hole
is more negative.\cite{LargeU} The hopping may therefore be 
enhanced by a somewhat larger factor than $\sqrt{N}$ in the large
$U$ case for certain systems.\cite{LargeU} 

In Fig. \ref{fig0}b we show schematically a doped system. The electron
added to the central site in an inverse photoemission process sees
a neighboring site which also has an extra electron. This additional
electron blocks one hopping channel. In the large $U$ case the 
effect is, however, much more dramatic, since hopping to a site with
an extra electron is then completely suppressed.

The doping of the system has a strong effect on the satellite structure.
Let us consider the large $U$ limit for a system 
with the average number $K+x$ electrons per site.
Then a fraction $1-x$ of the sites have $K$ electrons
and a fraction $x$ have $K+1$ electrons. If an electron is now removed
from a site with $K$ electrons, there is a strong coupling to final
states with $K-1$ electrons on that site. The energy of such a state 
is about $U$ higher than for a final state where all sites have $K$
or $K+1$ electrons. The result is a satellite at about the energy
$U$ below the main peak.\cite{Huefner} The weight of this satellite is 
reduced as $1-x$ is reduced.

\noindent
\begin{minipage}{3.375in}
\begin{figure}
\rotatebox{270}{
  {\epsfxsize=0.92in    \epsffile{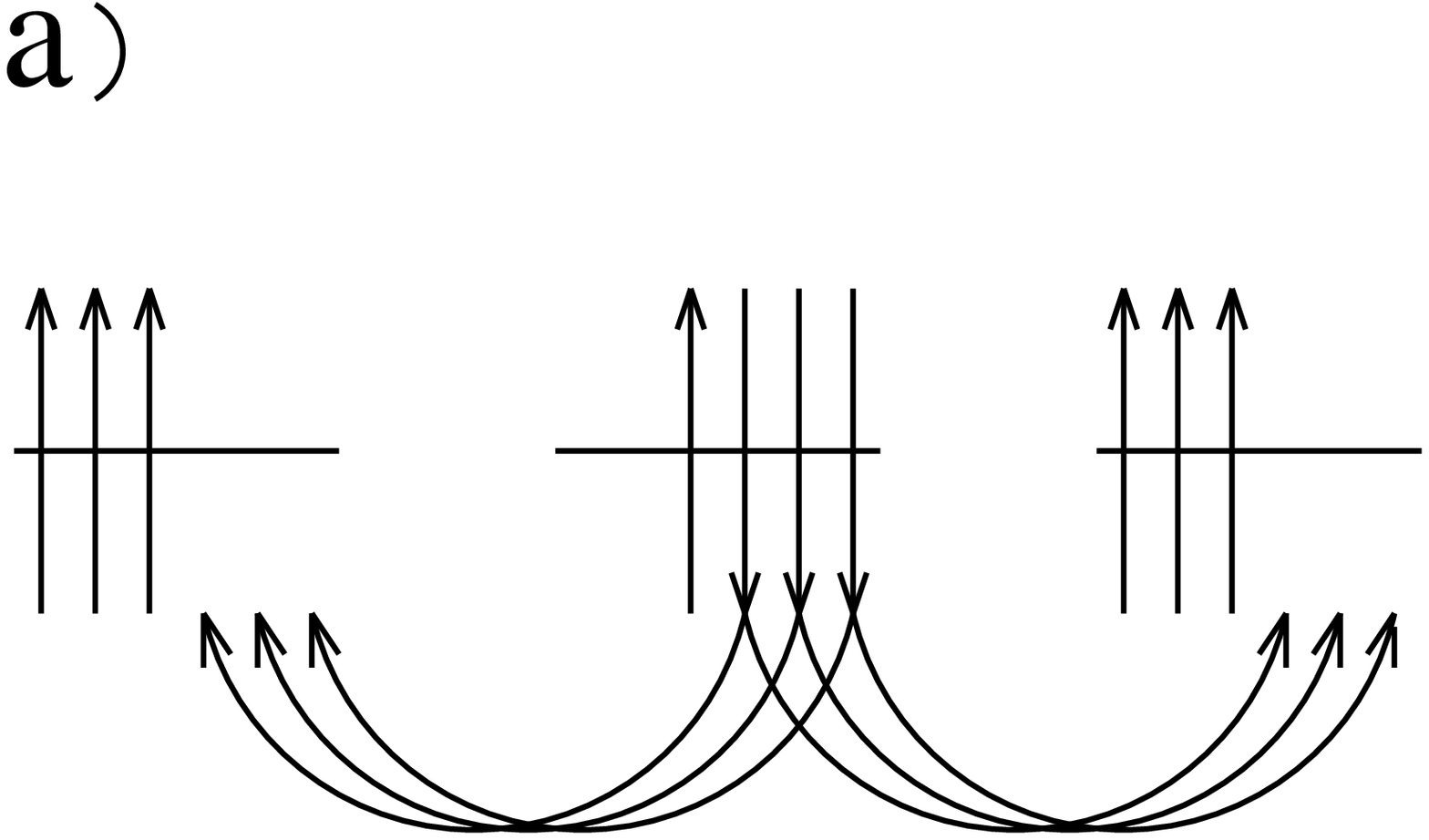}}} \hskip0.4cm 
\rotatebox{270}{
  {\epsfxsize=0.92in    \epsffile{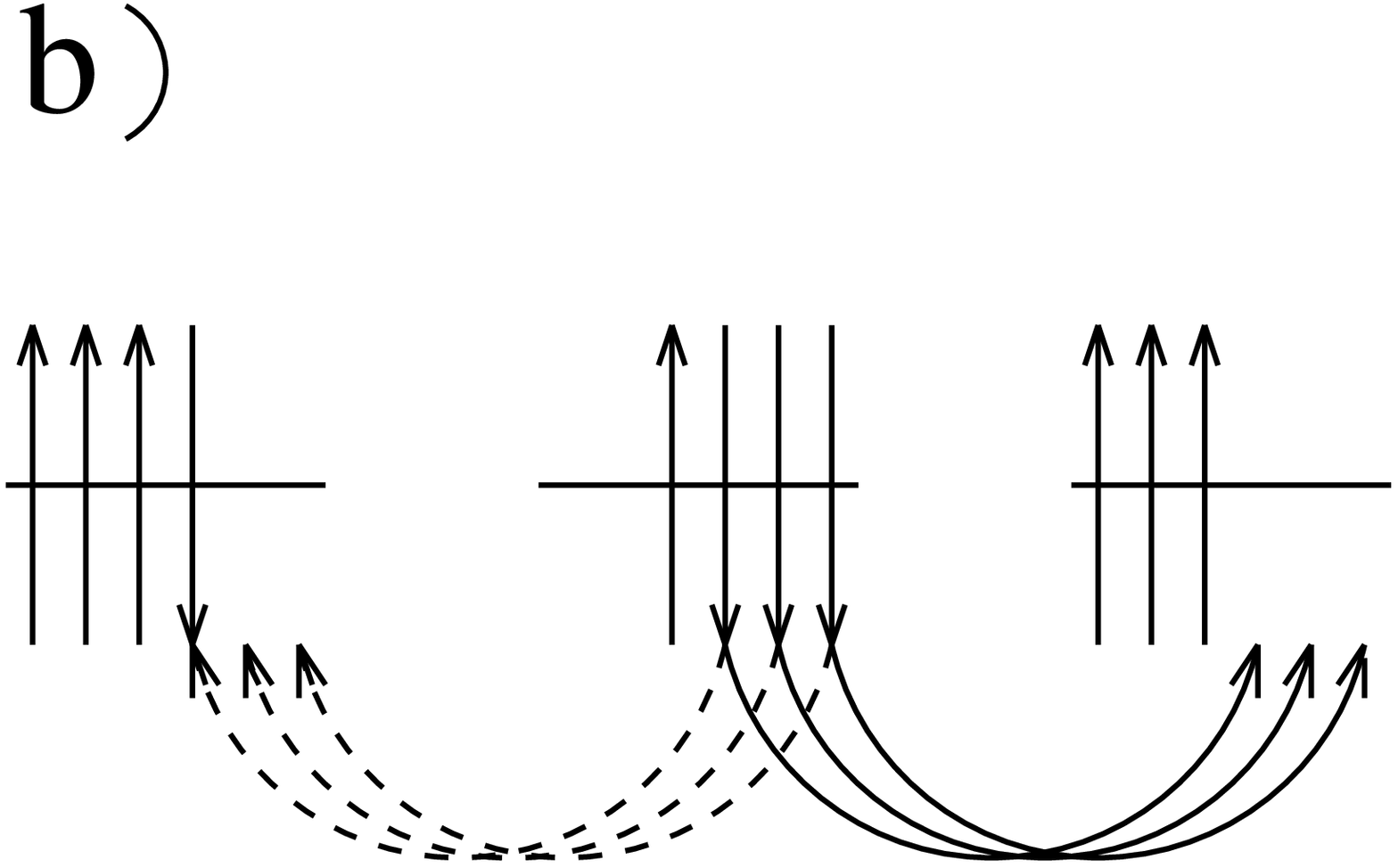}}}
  \hspace{4ex}
  \caption[]{\label{fig0}
  Hopping possibilities in the undoped (a) and doped (b) system.
  In the undoped system the extra occupancy can hop in  $N$ ways  to
  neighboring sites. In the doped case hopping to other sites with
  an extra occupancy is blocked if the Coulomb interaction is large. }
\end{figure}
\vspace{1ex}
\end{minipage}

\noindent
\begin{minipage}{3.375in}
\begin{figure}
  \centerline{\epsfxsize=2.900in \epsffile{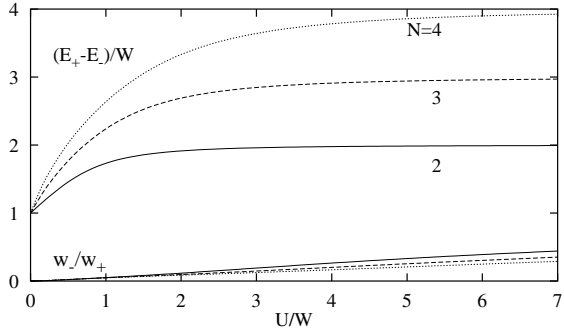}}
  \hspace{2ex}
  \caption[]{\label{fig1}Photoemission from a diatomic molecule with
orbital degeneracy $N$ and an   integer ($N$) number of electrons
per site at the ``Brillouin zone centre'' (+) and ``boundary'' (-). 
The dispersion (band width) $E_+-E_-$ and the ratio $w_-/w_+$ of 
the weights of the two peaks are shown. $W$ is the splitting 
(band width) between the bonding and antibonding levels for the 
noninteracting system.  The figure illustrates how the dispersion 
increases with    the Coulomb interaction $U$ and 
becomes proportional to $N$ for large $U$.}
\end{figure}
\vspace{1ex}
\end{minipage}

\subsection{Diatomic molecule}

We first consider the case of a diatomic molecule with only two 
sites and with two electrons. 
This model has been treated by Harris and Lange\cite{Harris}
in the absence of orbital degeneracy.
An electron is removed from a 
bonding or an antibonding orbital for the orbital quantum
number $m$ and  $\sigma=\uparrow$,
\begin{equation}\label{eq:5}
\psi_{\pm m \uparrow}={1\over \sqrt{2}}(\psi_{1m\uparrow}\pm\psi_{2m\uparrow}),
\end{equation}
and the result is averaged over $m$.
For the case $N=1$ the problem can easily be solved analytically,
and we find
\begin{equation}\label{eq:6}
\rho_{\pm}(\varepsilon)=\alpha_{\pm}\delta(\varepsilon-E_0(2)\pm t),
\end{equation}
where $\alpha_{\pm}=(1+x^2\pm\sqrt{1+x^2})/\lbrack 2(1+x^2)\rbrack$
and $E_0(2)=2|t|(x-\sqrt{1+x^2})$, with $x=U/(2|t|)$.
For $U\to 0$ the weight $\alpha_-$ of the antibonding state 
goes to zero, and for $U=0$ the spectrum has only one peak.  
For $U>0$ the separation between the peaks for $\rho_+$ and $\rho_-$ 
is $2|t|=W$, where $W$ is the 
one-particle band width. Thus the width of the angular
integrated PES spectrum is the full one-particle band width 
$W$,\cite{Brinkman} and not the width of the occupied part.
We can think of this width as a dispersional width, since the two
peaks in the angular integrated spectrum correspond to two different
values of $k$. In the solid, these $k$ values correspond to the 
Brillouin zone centre and boundary, respectively.

We next consider the two-site model for an orbital degeneracy
$N>1$. We first consider the case with $2N$ electrons, i.e., an
integer number of electrons per site. For the model  
(Eq.(\ref{eq:4})) of the hopping used 
here, the one-particle band width remains $W=2|t|$ independently
of $N$. Fig. \ref{fig1} shows the splitting  $E_+-E_-$ of the peaks 
in $\rho_{\pm}(\varepsilon)$ as a function of $U$ for different
values of $N$. The splitting grows with U from its noninteracting
value $W$ to its large $U$-limit $NW$.  
In the large $U$-limit the weights $w_\pm$ of the peaks in
$\rho_{\pm}$ become equal. As shown in the Fig. \ref{fig1}
this limit is reached quite slowly as $U/W$ grows.  
In addition to these peaks, there are satellites
at about $2U$ lower energy. These satellites have,
however, very small weights.
The reason for the large splitting between the bonding and 
antibonding peak is the degeneracy $N$. In the case of a two-site
model it can be shown\cite{LargeU} that the hopping is enhanced by a factor
$N$ instead of the factor $\sqrt{N}$ in the simple argument 
above.

\noindent
\begin{minipage}{3.375in}
\begin{figure}
  \centerline{\epsfxsize=2.900in \epsffile{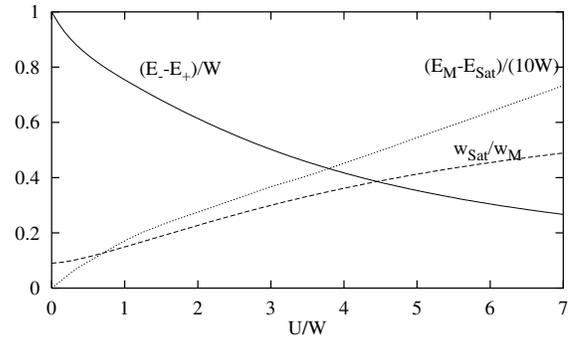}}
  \hspace{2ex}
  \caption[]{\label{fig2}Photoemission from a diatomic molecule with
orbital degeneracy $N=2$ and a noninteger (2.5) number of electrons 
per site. The dispersion (band width) $E_--E_+$, the satellite to main
peak splitting $E_M-E_{Sat}$ and the ratio of the satellite and main peak
weights $w_{Sat}/w_M$ are shown. $W$ is the splitting (band width) between 
the bonding and antibonding levels for the noninteracting system.
The figure illustrates how the dispersion is reduced as the Coulomb
interaction $U$ is increased.}
\end{figure}
\vspace{1ex}
\end{minipage}

We next consider the two-site model with $2N+1$ electrons, i.e.,
a noninteger number ($N+{1\over 2}$) of electrons per site. 
In the large $U$-limit, the ground-state now has $N+1$ electrons 
on one site and $N$ electrons on the other site. 
If an electron is removed from the site with $N+1$ electrons 
in a photoemission process, the system is left with $N$ electrons 
on each site. As discussed above, hopping is now completely suppressed
in the large $U$ limit,
and we expect a negligible dispersion in the
final state. If, on the other hand, the electron is removed from
the site with $N$ electrons, the result is a satellite, as discussed above.
These results are illustrated
in Fig. \ref{fig2} for the case $N=2$. For $U=0$ the electron
can be removed from a bonding ($k=+$) or antibonding ($k=-$)
orbital, giving a dispersion $W$. In this limit $\rho_{\pm}$
have just one peak each.
As $U$ is increased, the dispersion (separation between the
main peaks $E_{\pm}$) decreases and goes to zero for $U\to \infty$.
At the same time satellites develop at an energy about $U$ below 
the main peaks. The weights of these satellites approach the weight
of the main peaks in the large $U$-limit.
Since there are several satellites with slightly different energies,
we have calculated the average energy of the satellites in the  
two spectra $\rho_{\pm}$. The splitting between the main peak
and the average satellite position was then averaged for the two
spectra using the total satellite weight in each spectra as a weight 
factor.

For the diatomic molecule with $N>1$ we thus find that the dispersional
band width increases with $U$ for integer occupancy $(N)$ but decreases
with $U$ for half integer occupancy $(N+1/2)$ relatively to the noninteracting
width $W$.

\subsection{Linear chain}

The one-dimensional Hubbard model has been studied very
extensively in the literature.\cite{Shiba}  
We first consider a linear chain with six sites, periodic boundary 
conditions  and $N=1$. 
Fig. \ref{fig3}a shows the angular resolved spectrum for different 
values of $k$. The figure illustrates 
that there is a substantial dispersion as in the diatomic molecule. 
Thus the width of the occupied part of the spectrum is only $0.25W$
for noninteracting electrons, but for interacting electrons ($U/W=30$)
the width is about $W$\cite{Brinkman} for $N=1$.
The spectrum is more complicated 
than for the diatomic molecule, having several peaks for 
each $k$-value.  In the large $U$-limit, appropriate here, the spectrum
can be interpreted in terms of holons and spinons.\cite{Shiba}

Fig. \ref{fig3}b shows the angular resolved spectrum for $N=2$. 
The spectrum has more structures than for $N=1$ (Fig. \ref{fig3}a)
but the gross features are very similar. A major difference is,
however, that the energy scale is about a factor of two larger. 
Thus the width of the spectrum is now almost twice ($\sim 1.8$) the one-particle
band width $W$. This increased width is not due to the creation of
new satellites with a large binding energy, but due to a rather
uniform expansion of the energy scale for all the features present for $N=1$.
This is illustrated by the centre of gravity which has a stronger 
$k$-dependence for $N=2$. Thus we may think of the increased 
band width as to a large extent being due to an increased dispersion.
In this sense the situation is rather similar to the case of the diatomic
molecule considered before.

\noindent
\begin{minipage}{3.375in}
\begin{figure}
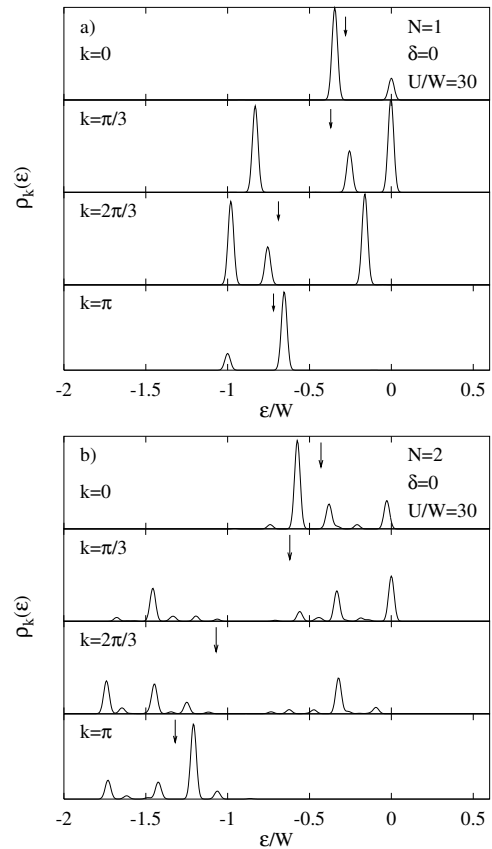

  {\epsfxsize=2.5in    \epsffile{anglinear1.epsi}} 
  \vskip0.2cm
  {\epsfxsize=2.5in    \epsffile{anglinear20.epsi}}
  \hspace{2ex}
  \caption[]{\label{fig3}Photoemission spectrum of a six atom linear
 chain with six electrons (integer occupancy=1, doping $\delta=0$) 
 as a function 
 of $k$ and $\varepsilon$. The orbital degeneracy is $N=1$ in a) and 
 $N=2$ in b). 
 The one-particle band width is $W$ and  $U/W=30$ is very large.
 The arrows show the centre of gravity of the main part of the spectrum,
 excluding the contribution from any satellite at about $\varepsilon/W
 \sim -30$ or lower. For plotting reasons a Gaussian broadening of 0.04$W$ 
(full width half maximum)
has been introduced. The energy zero is given by the leading peak.
Comparison of a) and b) shows that
the energy scale of the spectrum has been expanded by almost a factor
of two due to the degeneracy.}
\end{figure}
\vspace{1ex}
\end{minipage}

\noindent
\begin{minipage}{3.375in}
\begin{figure}
  \vskip1.5cm
  \centerline{\epsfxsize=2.5in \epsffile{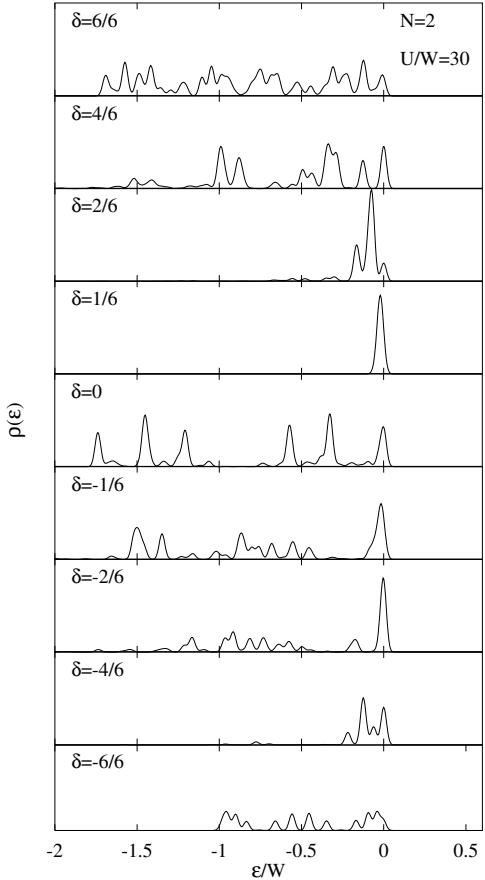}}
  \hspace{2ex}
  \caption[]{\label{fig5}Photoemission spectrum of a six atom linear 
chain as a function of doping $\delta$, where the number of electrons
is $(2+\delta)6$. The orbital degeneracy is $N=2$ and the one-particle 
band width is $W$. The figure illustrates how the width of the spectrum
is reduced when $\delta$ is changed  from 4/6 to 1/6 or from 0 to -4/6.
Observe the dramatic difference between doping $\delta=1/6$ and
$\delta=-1/6$. For integer fillings 1 ($\delta=-6/6$) and 2 ($\delta=0$)
and 3 ($\delta=6/6$) the spectrum is broad.}                                          
\end{figure}
\vspace{1ex}
\end{minipage}

We next consider the results for a doped system with $(2+\delta)6$
electrons and $N=2$. The angular integrated spectra are shown in 
Fig. \ref{fig5}. As the doping $\delta$ becomes increasingly negative,
the spectrum becomes narrower. The same tendency is seen when $\delta$
is positive and reduced from $4/6$ to $1/6$. For $\delta=1/6$,
 the width of the 
spectrum is essentially  zero (apart from an artifical 
broadening used for plotting
reasons). The explanation
is the same as for the diatomic molecule. With doping $\delta=1/6$
the system has two electrons on five sites and three electrons on one site. 
To obtain a contribution to the 
main peak the electron must be removed from the site with three  electrons. 
We then obtain a final state with two electrons on each site. Since      
hopping then costs the energy $U$, it is strongly suppressed and the width
of the spectrum goes to zero. 
This general tendency also shows up for 
intermediate dopings, and it reflects the reduced hopping possibilities 
for the photoemission hole if it is surrounded by other holes.
It is instructive to notice the asymmetry between doping $\delta=1/6$
and $\delta=-1/6$. In the first case hopping is completely suppressed
in the final states corresponding to the leading peaks,
as discussed above. In the case $\delta=-1/6$, on the other hand,
the probability that the photoemission hole can hop to a neighboring
site would only be reduced from 1 to 5/6, if the occupancies were spatially
uncorrelated, and there is a small reduction in the width. 
In inverse photoemission the situation is reversed, with a large
reduction in the width for $\delta=-1/6$ and a small reduction
for $\delta=1/6$.

\noindent
\begin{minipage}{3.375in}
\begin{figure}
  \centerline{\epsfxsize=2.500in \epsffile{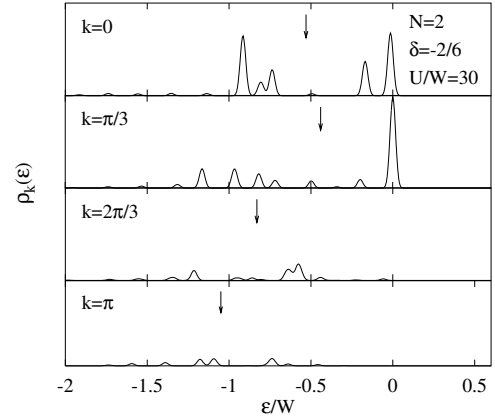}}
  \hspace{2ex}
  \caption[]{\label{fig6} The same as in Fig. \ref{fig3}b but for the  
 doping $\delta=-2/6$ (ten electrons). 
 Comparison with $\delta=0$ (Fig. \ref{fig3}b) shows that the spectra with the 
 lowest centre of gravity ($k=2\pi/3$, $\pi$) have lost  weight 
 and that the centre of gravity of each spectra tends to move upwards.}
\end{figure}
\vspace{1ex}
\end{minipage}

To see how this happens in more detail, we show in Fig. \ref{fig6}
the angular resolved spectrum for $\delta=-2/6$. Compared with the 
$\delta=0$ results in Fig. \ref{fig3}b, the spectrum for $k=\pi$ has 
a small weight. Since the centre of gravity of this spectrum is low,
the reduction of its weight leads to a reduced width of the 
angular integrated spectrum. The same applies to a smaller extent to 
the $k=2\pi/3$ spectrum. Furthermore, the centres of gravity of
the spectra for $k=\pi/3$ and $2\pi/3$ have moved upwards. This
can be viewed as a reduction of the dispersion, and it also contributes 
to the reduced width of the angular integrated spectrum. This upward 
movement of the centre of gravity is partly due to an upward movement 
of certain peaks and partly due to a reduced weight of peaks with
large binding energies.

Fig. \ref{fig7} shows the angular resolved spectrum for a positive 
doping. In this case the centres of gravity have moved up strongly for 
$k=0, \pi/3$ and $2\pi/3$. 
In addition, the spectrum for $k=\pi$ now has a negligible
weight, while for $\delta=0$ it contributed substantially to the 
width of the spectrum due to its low centre of gravity. 
In this case the reduction of the width of the angular integrated spectrum
is, however, mainly due to a reduced dispersion, while the situation
is less well-defined in Fig. \ref{fig6}.

\noindent
\begin{minipage}{3.375in}
\begin{figure}
  \centerline{\epsfxsize=2.500in \epsffile{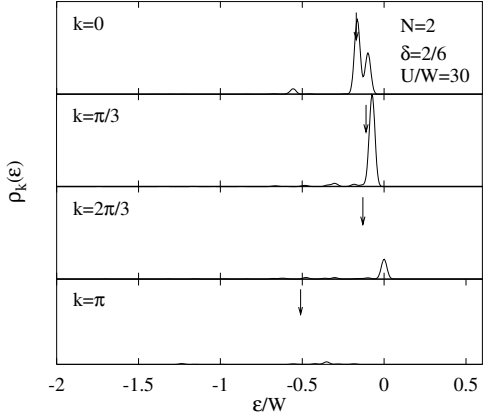}}
  \hspace{2ex}
  \caption[]{\label{fig7} The same as in Fig. \ref{fig3}b but for the  
 doping $\delta=2/6$ (fourteen electrons). 
 Comparison with $\delta=0$ (Fig. \ref{fig3}b) shows
 that the centres of gravity of the spectra for $k=0, 
\pi/3$ and  $2\pi/3$ have moved upwards, strongly reducing the dispersion.
 Furthermore, the spectrum with the 
 lowest centre of gravity ($k=2\pi/3$, $\pi$) has  lost most of its weight.}  
\end{figure}
\vspace{1ex}
\end{minipage}

It is also interesting to consider integer filling away from half-filling.
The arguments in Sec. II.B can be straightforwardly generalized  
to any integer filling.\cite{filling}
If the filling is $K\le N$, there are $K$ electrons which can fill a hole
on the site where a hole was created in the photoemission process. 
If the filling is $K>N$ there are $2N-(K-1)$
holes which can be filled on the site where a hole was created in 
the photoemission process. 
Above we found that the width of the photoemission spectrum
grows linearly with the number of hopping possibilities for a linear chain.  
For the fillings $K=1$ and $K=3$ this suggests the band 
widths $W$ and $2W$, respectively, for $N=2$. These results are 
confirmed by the explicit calculations shown in Fig. \ref{fig5}.
In inverse photoemission we expect $K+1$ hopping possibilities 
for $K<N$ and $2N-K$ hopping possibilities for $K\ge N$.

Up to now the linear chain has been discussed in the limit of $U/W>>1$.
Fig. \ref{fig8} illustrates the validity of these conclusions for intermediate
values of $U/W$. The figure shows the width                         
\begin{equation}\label{eq:7}
{\mathcal {W}}=2\sqrt{S_2-S1^2},
\end{equation}
where $S_i$ is the $i$th  moment of the spectrum. Also shown is the 
centre of gravity relative to the leading peak.  
The figure illustrates that the spectrum is broadened even for
small values of $U$. For $U/W> 1$ the degeneracy furthermore becomes 
important and the width is substantially larger for $N=2$ than $N=1$.

Fig. \ref{fig8a} shows the spectra for $U/W=1.5$ for $N=1$ (a) 
and $N=2$ (b). The figure illustrates that a substantial part of 
the large width in the $N=2$ case comes from the spectra
for $k=2\pi/3$ and $k=\pi$, which have small weights
but structures at large binding energies. Such contributions
would probably be very hard to detect experimentally. There
is, however, also a downward shift in the spectra for $k=0$ 
and $k=\pi/3$.

\noindent
\begin{minipage}{3.375in}
\begin{figure}
  \centerline{\epsfxsize=2.500in \epsffile{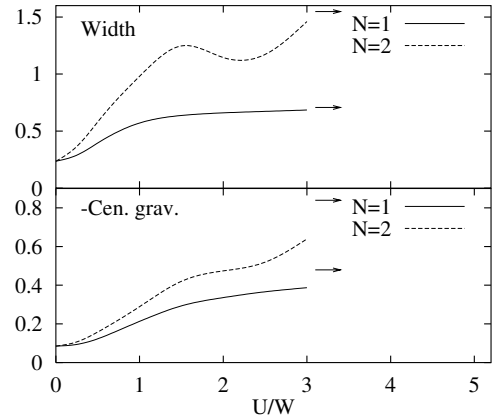}}
  \hspace{2ex}
  \caption[]{\label{fig8} Width and the negative of the 
  centre of gravity (relative to 
  the leading peak) of the spectrum for $N=1$ and $N=2$ as a 
  function of $U/W$. The horizontal arrows show the results 
  for $U/W=30$. 
The figure illustrates how the effects discussed
  remain important down to $U/W\sim 1$.  }
\end{figure}
\vspace{1ex}
\end{minipage}

\noindent
\begin{minipage}{3.375in}
\begin{figure}
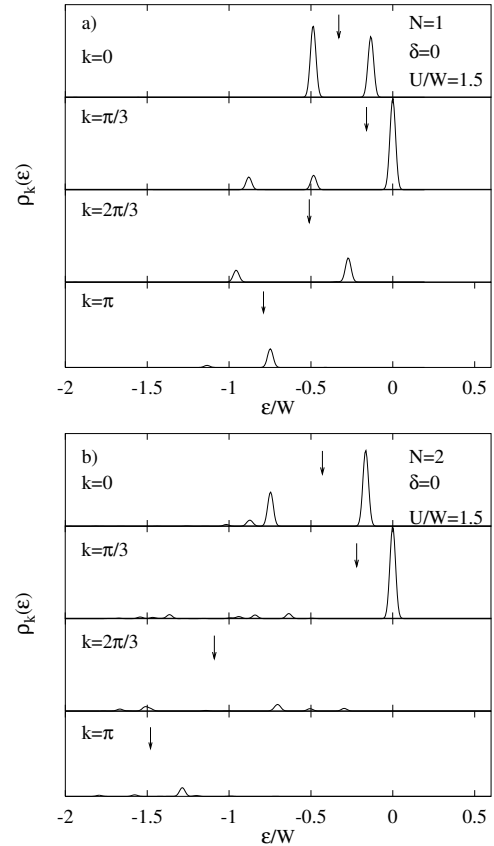

  \centerline{\epsfxsize=2.500in \epsffile{anglinear1u15.epsi}}
  \vskip0.2cm
  \centerline{\epsfxsize=2.500in \epsffile{anglinear20u15.epsi}}
  \hspace{2ex}
  \caption[]{\label{fig8a}Same as in Fig. \ref{fig3} but for          
  $U/W=1.5$. The figure illustrates that the effects in Fig. \ref{fig3}
  are substantially reduced at intermediate values of $U$, but that
  some effects still remain.  }
\end{figure}
\vspace{1ex}
\end{minipage}

\noindent
\begin{minipage}{3.375in}
\begin{figure}
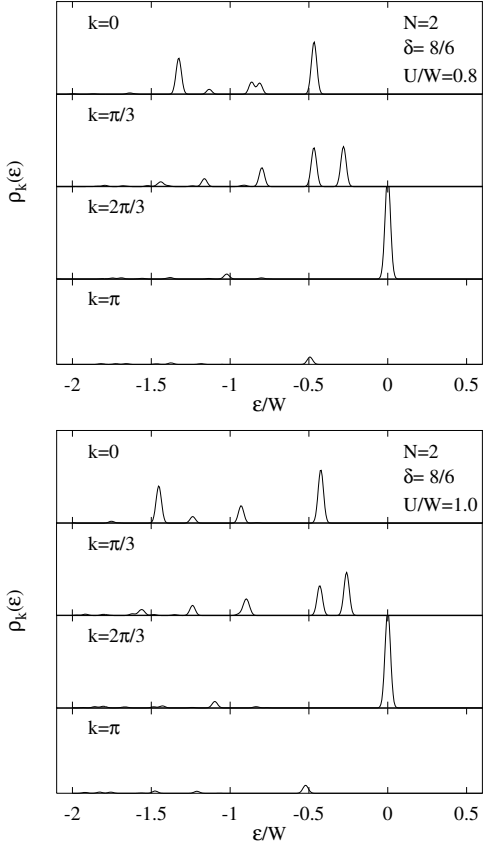

  \centerline{\epsfxsize=2.500in \epsffile{anglinear2p8u08.epsi}}
  \vskip0.2cm
  \centerline{\epsfxsize=2.500in \epsffile{anglinear2p8u10.epsi}}
  \hspace{2ex}
  \caption[]{\label{fig8b}The same as in Fig. \ref{fig3}b but for     
  doping $\delta=8/6$ (0.67 holes per site) and $U/W=0.8$ (a) and
  $U/W=1.0$ (b). These parameters may be typical for Ni metal. 
  Comparison of (a) and (b) illustrates how the peaks above 
  $\sim-W/2$ move upwards as $U$ is increased (``band narrowing'')
  while the peaks below $\sim -W$ move downwards (``satellites''). 
  For $U=0$ the width of the photoemission spectrum is $0.75W$.  }
\end{figure}
\vspace{1ex}
\end{minipage}

Finally we consider parameters which may be considered as representative
for Ni metal. Thus we have chosen $U/W=0.8$ (Fig.~\ref{fig8b}a)
and $U/W=1.0$ (Fig.~\ref{fig8b}b) and the doping $\delta=8/6$,
which corresponds to 0.67 holes per site. According to Auger measurements,
$U/W$ for Ni is close to but slightly smaller than one.\cite{Antonides} 
The number of $3d$-holes in Ni  is about 0.6.
As discussed earlier, because of the noninteger number of electrons    
per sites, we expect satellites that move down in energy with $U$.
To distinguish between these types of satellites
and the ``main'' band,  we have performed the calculation for two 
values of $U$. In Fig.~\ref{fig8b} the peaks below about $-W/2$ move
downwards as $U$ is increased, while the peaks above $-W/2$ move
upwards. Although $U/W$ is not large in this case, we can therefore
nevertheless identify the peaks below $-W/2$ as satellites. This
means that the main band has only a width of about $0.45W$. This     
should be compared with the width $0.75W$ of the the occupied part 
of the band for $U=0$. This means that the band width has been reduced 
by about 40 $\%$. This is a bit larger than the experimental result
($25 \%$), but it illustrates that the model gives a qualitatively
correct band width, in spite of its extreme simplicity.

To understand the band narrowing, we observe that for a site with
two electrons the probability that a given neighboring site also              
has two electrons would be 1/6 if the sites with two electrons 
were uncorrelated. There is then one hopping possibility for the hole
to such a site, while hopping to all other site costs the energy $U$. 
This would then suggest a reduction of the band width to $W/6$.  
Actual calculations for $U/W=30$ give reduction by a factor somewhat
smaller than five. The much less dramatic reduction of the band width
in Fig. \ref{fig8b} is due to the value of $U$ being intermediate.

\noindent
\begin{minipage}{3.375in}
\begin{figure}
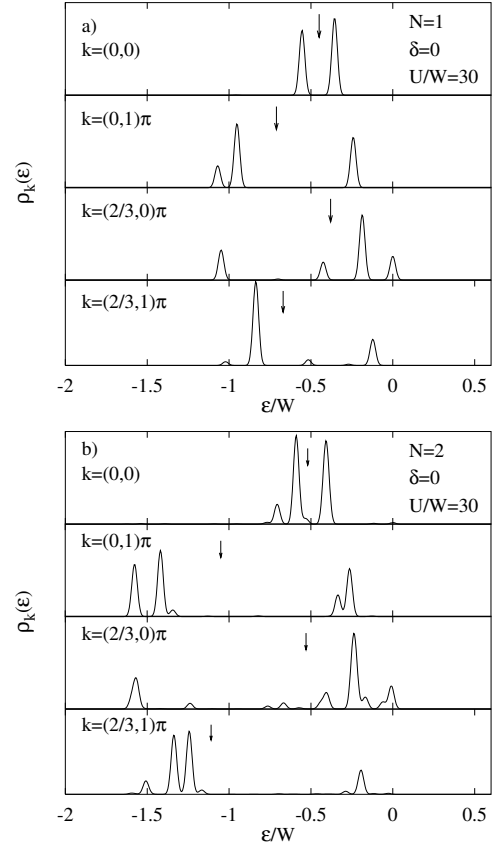

  \centerline{\epsfxsize=2.500in \epsffile{sixcluster10.epsi}}
\vskip0.2cm
  \centerline{\epsfxsize=2.500in \epsffile{sixcluster20.epsi}}
  \hspace{2ex}
  \caption[]{\label{fig9} Photoemission spectrum of a cluster with
six atoms in a rectangular arrangement ($2\times 3$) with six electrons
(integer occupancy) and $U/W=30$ for $N=1$ a) and $N=2$ b). 
 The arrows show the centre of gravity of the main part of the spectrum,
 excluding any satellite at $\varepsilon/W \sim -30$ or lower.
 Comparison of a) and b) shows how the energy scale has been 
expanded by about a factor of 1.5 for $N=2$ compared with $N=1$. }
\end{figure}
\vspace{1ex}
\end{minipage}

\subsection{Six atom ''rectangular''  cluster}
As an further example we study a cluster with six atoms arranged in 
a rectangle with three times two sites, using periodic boundary conditions.
Fig. \ref{fig9}a shows the spectrum for the orbital degeneracy $N=1$.
In this case the width of the photoemission spectrum is even larger than
the one-particle band width $W$. The reason\cite{LargeU} is the 
frustration in the six atom cluster due to the periodic boundary 
conditions. Because of this, the one-particle band width $W=5|t|$ is smaller
than one might have expected ($6|t|$) from the hopping integral and the number 
of nearest neighbors (=3). The frustration effect on the many-body 
spectrum for a large $U/W$ is smaller, giving a width of the spectrum
of about $5.75|t|=1.15W$ already for $N=1$. The width of the angular 
integrated spectrum is to a substantial amount due to the dispersion.

Fig. \ref{fig9}b shows the spectrum for $N=2$. The spectrum shows large 
similarities with the case $N=1$ (Fig. \ref{fig9}a), except that the 
energy scale is larger by a factor of about 1.5. This behavior lies
in between $\sqrt{N}$ and $N$ and has been discussed in
 Ref. \onlinecite{LargeU}. Fig. \ref{fig9} illustrates that 
the increase of the width of the spectrum with $N$ is not limited 
to the linear chain but a general phenomenon. 

\subsection{A$_3$C$_{60}$ (A= K, Rb)}

The photoemission from A$_3$C$_{60}$ (A=K, Rb) is interesting, since
the experimental width of the spectrum is about 1.5 eV,\cite{Chen} 
while the occupied width of the one-particle spectrum is only of the
order of 0.3 eV.\cite{Satpathy}
As discussed in the introduction, however, the experimental 
spectrum is angular resolved, and it is not known if this large
width is due to dispersion or satellites.
We have considered multi-band Hubbard model in Eq. (\ref{eq:1})
to describe the C$_{60}$ $t_{1u}$ level with    
degeneracy three. The hopping matrix elements are obtained from
a tight-binding theory,\cite{Orientation,Satpathy,MazinAF}
which takes the orientational disorder in A$_3$C$_{60}$ into account.
We use parameters which give a one-particle full band width $W=0.6$ eV.
The angular integrated photoemission spectrum is shown in Fig. \ref{fig11}.
The total width of the spectrum is about 1.5 eV. This width is larger
than $\sqrt{N}W\sim 1$ eV. This may be due to the width growing
somewhat faster than $\sqrt{N}$ and to the system being frustrated 
because of the fcc lattice.

The calculated width of the spectrum is of the same order as observed 
experimentally. In contrast to the experimental spectrum there is,
however, no weight in the range -0.3 to -0.7 eV, while the experimental
 spectrum is smoothly reduced with increasing binding energy.
The calculated spectrum also has much too much weight at small 
binding energies.
The spectrum in Fig. \ref{fig11} can therefore not explain experiment.
The main contribution to the broad experimental spectrum is therefore
probably from phonon and plasmon satellites, 
as suggested earlier.\cite{Knupfer} The degeneracy 
mechanism discussed here should, however, make some contribution to 
the weight at large binding energies.\cite{Antropov}

\noindent
\begin{minipage}{3.375in}
\begin{figure}
  \centerline{\epsfxsize=2.500in \epsffile{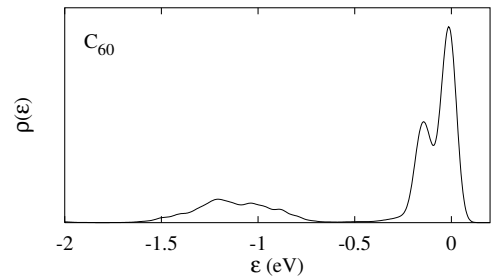}}
  \hspace{2ex}
  \caption[]{\label{fig11} Photoemission spectrum from a model of   
four C$_{60}$ molecules for $U=1.3$ eV. The figure illustrates that 
apart from the main peak there is additional weight at about -1 to 
-1.5 eV. This weight
does not move as $U$ is increased and it is related   to the three-fold
orbital degeneracy.}
\end{figure}
\vspace{1ex}
\end{minipage}

\noindent
\begin{minipage}{3.375in}
\begin{figure}
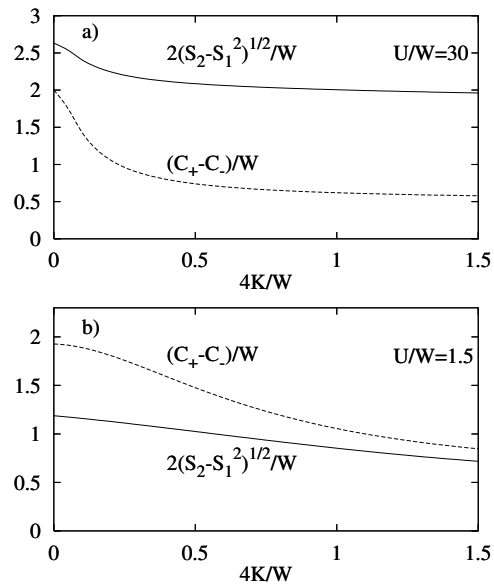

  \centerline{\epsfxsize=2.500in \epsffile{widthmult1.epsi}}
\vskip0.2cm
  \centerline{\epsfxsize=2.500in \epsffile{widthmult2.epsi}}
  \hspace{2ex}
  \caption[]{\label{fig9a} 
  The dispersion $C_+-C_-$ and the width $2\sqrt{S_2-S_1^2}$ of 
the photoemission
 spectrum as a function of the multiplet splitting $4K$ for a two-site
system, where $K$ is the multiplet integral. $C_{\pm}$ is the centre 
of gravity for $k=\pm$. A large 
($U/W=30$) and an intermediate ($U/W=1.5$) value of $U$ were considered. 

}
\end{figure}
\vspace{1ex}
\end{minipage}

\subsection{Multiplet effects}

An interesting question is how these results are influenced by 
multiplet effects. If a site with an extra electron or hole
is in a given multiplet state, the hopping of the extra occupancy
to a neighboring site with a similar energy may be restricted,
because the multiplet effects lift degeneracies.

Here we consider a model, where the multiplet effects  are described 
by the exchange integral $K$ between two different orbitals and 
the difference  $\delta U\equiv U_{xx}-U_{xy}$ between the direct
Coulomb integral for equal and unequal orbitals. Here we use 
$\delta U=2K$. Thus we add the multiplet terms 
\begin{eqnarray}\label{eq:m1}
H_U=&&{2 \over 3}\delta U\sum_{im} n_{im\uparrow}n_{im\downarrow}
-{1\over 3}\delta U\sum_{i\sigma\sigma^{'}}\sum_{m<   m^{'}}n_{i\sigma m}
n_{i\sigma^{'}m^{'}}  \nonumber  \\
+&&{1\over 2}K \sum_{i\sigma\sigma^{'}}\sum_{m\ne m^{'}}
\psi^{\dagger}_{i\sigma m}\psi^{\dagger}_{i\sigma^{'} m^{'}}
\psi_{i\sigma^{'} m}\psi_{i\sigma m^{'}}   \\
+&&{1\over 2}K \sum_{i\sigma}\sum_{m\ne m^{'}}
\psi^{\dagger}_{i\sigma m}\psi^{\dagger}_{i-\sigma m}
\psi_{i-\sigma m^{'}}\psi_{i\sigma m^{'}}  \nonumber  
\end{eqnarray}
to the Hamiltonian in Eq.~(\ref{eq:1}).
This part of the Hamiltonian has multiplets with the energies
0, $2K$ and $4K$ relative to the lowest multiplet, i.e., a splitting
of $4K$.

We consider a system with two sites,    the orbital degeneracy $N=2$ 
and four electrons, i.e., integer occupancy.
Fig. \ref{fig9a} shows results for the width $2\sqrt{S_2-S_1^2}$ and the 
separation $(C_+-C_-)$ in the centres of gravity of the spectrum, where $S_i$
is the $i$th  moment.  
For a large $U$ (Fig. \ref{fig9a}a) the width of the spectrum is gradually
reduced with the multiplet splitting as expected, but the reduction is 
moderate.  The "dispersion" measured through $C_+-C_-$ drops is reduced 
much more. The reason
for this drop is a change in the character of the spectrum. 
For $K=0$ the spectra for $k=\pm$ have essentially just one peak each.
As $K$ is increased, new satellites are formed on the energy scale $W$
(not $K$!), which lead to rapid reduction in the difference $C_+-C_-$,
without changing the total width of the spectrum drastically.
The creation of new satellites are related to changes in the ground-state
wave function. Due to the integer occupancy in the ground-state,
the hopping energy is very small, and the energy scale over which
the multiplet effects become important is very small. Fig. \ref{fig9a}b
shows results for an intermediate value of $U$. In this case the hopping  
in the initial state is more important, and a small value of $K$ has a
small influence on the ground-state. Both $C_+-C_-$ and $2\sqrt{S_2-S_1^2}$
are now reduced with $4K$ at a much slower rate.  
Due to the smaller value of $U$, however, both quantities are smaller
for $K=0$ than in Fig.~\ref{fig9a}a. The results above suggest that 
multiplet couplings reduce the effects discussed in this paper, but 
that these effects are still present for intermediate multiplet splittings.

\section{Specific heat}

If the dispersion of the states is indeed increased due to the 
degeneracy, this should also show up as a reduction in the 
specific heat.  
We thus  consider the specific heat
\begin{equation}\label{eq:7a}
c_v={ d E(T)\over dT},
\end{equation}
where $E(T)$ is the energy of the system as a function of 
the temperature $T$. To perform the calculation at finite 
$T$, we use a technique developed by Jaklic and 
Prelovsek,\cite{Prelovsek} and based on the Lanczos method.
In this technique a certain number of states $N_0$ are chosen 
randomly, and for each state $|n\rangle$ the appropriate expectation 
values $\langle n| {\rm exp}(-H/T) |n\rangle$ and 
 $\langle n| {\rm exp}(-H/T)H |n\rangle$ are calculated using
the Lanczos method. We here use the grand canonical ensemble,
i.e., the number of electrons is not fixed.
In this approach the specific heat is independent of the degeneracy
for noninteracting electrons, as it should. The use of the canonical 
ensemble leads to a moderate degeneracy dependence even for $U=0$,
due to the finite size of the system and the change of the number of 
electrons with degeneracy, i.e., the system being closer to infinite
for a larger $N$.

We consider the Hubbard model for a six-atom cluster with the atoms 
arranged in a $3\times 2$ rectangle, as above. The        
hopping integrals in Eq. (\ref{eq:4}) are used. 
In the large $U$-limit the total energy at half-filling for a
system with $M$ sites and the orbital degeneracy $N$ is
\begin{equation}\label{eq:8}
E(T=0)={1\over 2}MN(N-1)U.
\end{equation}
Here we have neglected small terms of the order $-t^2/U$.
For a very large temperature we obtain
\begin{equation}\label{eq:9}
E(T=\infty)=({1\over 2})^2MN(2N-1)U.
\end{equation}
The energy increase per electron is then                         
\begin{equation}\label{eq:10}
{1\over MN}\lbrack E(T=\infty)-E(T=0)\rbrack={1\over 4}U,       
\end{equation}
i.e., independent of the degeneracy $N$. Integrating $c_v/(MN)$ over all 
temperatures should then give a quantity which is independent of 
$N$ for large $U$. In this          limit, we therefore expect
the specific heat to be reduced with $N$ but to     
remain appreciable over a correspondingly larger temperature range.

We may think of the contributions to the specific heat as resulting 
from the excitation of spin and charge degrees of freedom.
Due to its finite size, the system has a band gap even for small values 
of $U$. For $U/W=1.5$ and $N=1$ the gap is $E_g=0.76W$ and for 
$N=2$ we find $E_g=0.45W$. For $T$ much smaller than the band gap,
charge degrees of freedom are not excited, and only spin excitations 
contribute to the specific heat. Fig. \ref{fig12} shows the 
spin correlation function
\begin{equation}\label{eq:11} 
{1\over 4NM}\sum_{\langle i,j\rangle m}\langle \sigma_{im}^z 
\sigma_{jm}^z \rangle,  
\end{equation}
where the summation is over nearest neighbors and $\sigma^z_{im}$ is the
Pauli spin matrix $\sigma^z$ for orbital $m$ on site $i$.  
The figure illustrates
that the spins become uncorrelated over an energy scale of about
a few tenths of $W$. This energy scale is smaller than the band gap
and for the corresponding temperatures the spin excitations         
should dominate the specific heat. The specific heat is also shown in Fig.
\ref{fig12}.                          
The present system may be too small to describe the spin degrees of 
freedom of a large system, since at low temperatures only a very
small number of states contribute to the specific heat, and since 
we do not know if these few states are representative of a larger system. 
Thus the sharp peak in $c_v$ for small $T$'s may be a defect of the small size
of the system.
We are interested in metallic systems, i.e.,
systems where the band gap goes to zero with increasing system size.
For these systems the charge degrees of freedom should become 
increasingly important for low temperatures as the size of the system 
grows. Because of this and because of the uncertainty in the 
contribution from the spin degrees of freedom we here concentrate 
on the charge degrees of freedom, dominating for $T \gtrsim 0.2W$
for the six atom cluster.

For low temperatures in this range, the specific heat is smaller for 
$N=2$ than for $N=1$, in agreement with the expectations above.
The specific heat decreases more slowly
with $T$ for $N=2$, remaining appreciable up to larger values of $T$.
This is required to satisfy Eq.~(\ref{eq:11}).       
Finally we remark that although we have not discussed the spin degrees
of freedom further, these degrees give an important contribution
for many strongly correlated systems.

\section{Concluding remarks}

We have studied the effects of orbital degeneracy $N$ and doping
on the photoemission spectrum in the limit of a large Coulomb
interaction $U$.
We find that the degeneracy increases the dispersional width of the
spectrum for integer occupancy. Deviations from integer occupancy
tends to reduce the width of the spectrum, in particular if the 
occupation is slightly larger than integer. These effects are gradually 
reduced as $U$ is reduced or as multiplet effects are taken into account.
 We also find that the degeneracy tends to
reduce the contribution to the specific heat from the charge degrees
of freedom, as one would expect from the increased dispersion.

An interesting approach for treating correlation effects is the
dynamical mean field theory.\cite{Georges} In this approach the
self-energy is assumed to have no wave vector dependence, 
which is correct for a system with infinite 
dimension.\cite{Metzner,Muller}
The present results show that in a Hubbard model with integer occupancy,
 the orbital degeneracy increases 
the ${\bf k}$ dependence for intermediate to large values of 
the Coulomb interaction $U$. This suggests that the dimension 
where the dynamical mean field theory becomes accurate
increases with the orbital degeneracy. 

\noindent
\begin{minipage}{3.375in}
\begin{figure}
  \centerline{\epsfxsize=3.200in \epsffile{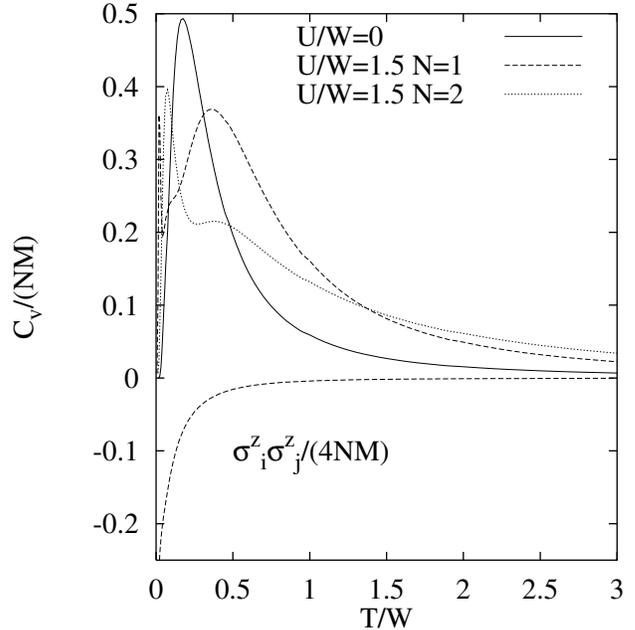}}
  \hspace{2ex}
  \caption[]{\label{fig12} Specific heat for a six-atom $3\times2$
  cluster with $U/W=1.5$ as a function of temperature $T$ and 
degeneracy $N$. The result for $U=0$ is also shown.
The spin correlation function $\langle \sigma_i^z\sigma_j^z\rangle$
(for $N=1$)
illustrates that the spin degrees of freedom contribute only for
 small $T$, while the charge degree of freedoms contribute for
larger $T$. The quantities have been normalized to the number of
electrons $NM$. }
\end{figure}
\vspace{1ex}
\end{minipage}

It should be emphasized that we have here studied a (degenerate)
one-band Hubbard model.
For each correlated system one has to consider whether or not 
this model is appropriate before applying the results above.
For certain systems, e.g, heavy fermion and High $T_c$ compounds, 
the interaction between different types of orbitals plays an
important role, and it would be interesting to study how this
influences the considerations above.

\section{Acknowledgements}
We would like to thank L.F. Feiner, P. Horsch, E. Koch, R.M. Martin 
A.M. Oles, and T.M. Rice for stimulating discussions.

\end{multicols}
\end{document}